\documentclass[aps,floatfix,superscriptaddress,notitlepage]{revtex4-1}
\usepackage{hyperref,amsfonts,amsmath,amssymb,graphics,graphicx,dcolumn,bm,multirow,
enumerate,verbatim,threeparttable,natbib}

\begin{document}
\title{A theoretical model for the associative nature of conference participation}
\author{Jelena Smiljani\'c}
 \affiliation{Scientific Computing Laboratory, Institute of Physics Belgrade, University of Belgrade, Pregrevica 118, 11080 Belgrade, Serbia}
\affiliation{School of Electrical Engineering, University of Belgrade, P.O. Box 35-54, 11120 Belgrade, Serbia}
 \author{Arnab Chatterjee}
  \affiliation{Condensed Matter Physics Division, Saha Institute of Nuclear Physics, 1/AF Bidhannagar, Kolkata 700064, India}
 \author{Tomi Kauppinen}
  \affiliation{Department of Computer Science, Aalto University School of Science, P.O. Box 11000, FI-00076 Aalto, Finland}

 \author{Marija Mitrovi\'c Dankulov}
 \email{mitrovic@ipb.ac.rs}
 \affiliation{Scientific Computing Laboratory, Institute of Physics Belgrade, University of Belgrade, Pregrevica 118, 11080 Belgrade, Serbia}
%
%

\begin{abstract}
Participation in conferences is an important part of every scientific career. Conferences provide an opportunity for a fast dissemination of latest results, discussion and exchange of ideas, and broadening of scientists' collaboration network. The decision to participate in a conference depends on several factors like the location, cost, popularity of keynote speakers, and the scientist’s association with the community. Here we discuss and formulate the problem of discovering how a scientist’s previous participation affects her/his future participations in the same conference series. We develop a stochastic model to examine scientists' participation patterns in conferences and compare our model with data from six conferences across various scientific fields and communities. Our model shows that the probability for a scientist to participate in a given conference series strongly depends on the balance between the number of participations and non-participations during his/her early connections with the community. An active participation in a conference series strengthens the scientist’s association with that particular conference community and thus increases the probability of future participations.
\end{abstract}

\maketitle

\section*{Introduction}
Social data at a large scale is nowadays available over the internet. Researchers are making the best use of these data to find trends, 
statistics and patterns, which sometime reveal as robust features, similar to `laws' in natural science. In recent years, 
a huge community of researchers~\cite{lazer2009life} including mathematicians, statisticians, computer scientists, theoretical physicists, sociologists, economists, financial analysts, geographers, anthropologists, and biologists of various sub-disciplines have contributed to a larger, developing field, commonly known as  `computational social science'~\cite{cioffi2010computational}. Empirical data, after a rigorous analysis produces information that is of immense interest for theoreticians. Statistical mechanics, which has been proved 
to be versatile in modeling phenomena across different areas of physics, and beyond, seems to be the most desired tool even for the above emerging discipline~\cite{castellano2009statistical,Sen2014sociophysics}.

The abundance of a new data about scientific activities such as publications, collaborations, and citations led to the emergence of a new interdisciplinary field of research about science and how science works \cite{schweitzer2014}. These studies provide insights about the impact of scientists and their publications \cite{radicchi2008,radicchi2009, wang2013}, authors' reputation and scientific success \cite{sinatra2014}, patterns of collaboration and their impact on authors' reputation \cite{petersen2012,azoulay2010}, the role of cumulative advantage in career longevity \cite{petersen2011,petersen2014b} and scientific mobility \cite{deville2014} among many other things. Despite the attention given to publication records and citation patterns, another integral part of modern science, scientific meetings, have so far been largely overlooked. This negligence is particularity interesting, given the pervasive role of the meetings in scientific disciplines. Scientific meetings provide arenas for a fast dissemination of the latest results, exchange and evaluation of ideas as well as a knowledge extension. 
However, the most important function of scientific meetings is to facilitate social contacts. They provide an opportunity and platform to extend the network of collaborators through the creation of new contacts, and to strengthen existing links by getting reacquainted with old friends.

Undoubtedly, conference participation has a very positive impact on scientific career. In addition to the opportunities they provide, attending a scientific meeting can be very costly, both in terms of time and money. Bearing in mind that the number of national and international meetings have drastically increased in the last few decades, it is clear that scientists are now pressed to make a careful selection of the meetings they will attend. Extensive studies \cite{borghans2010,mair2009,witt1995} have shown that conference characteristics, such as the attractiveness and the reachability of the location or the choice of keynote speakers affect the decision of  scientists to attend a meeting. The role of the social component in conference choice is so far unexplored, mainly due to lack of quality data. The social component, such as the association with a conference community or conference inclusiveness, are of crucial importance when it comes to whether a conference participation was beneficial or not. This is particularly evident in the case of young scientists, who are new to a community and struggle to overcome the social obstacle of an initial contact \cite{essay1,essay2}. One of the rare studies on conference participation \cite{vandijk2006} has shown that conferences have a stable core of regularly attending participants, regardless of the conference location and distance. Having in mind that characteristics like the attractiveness of a location and the quality of keynote speakers are fluctuating from one year to another, it is clear that social component of a  conference strongly influence the scientists decision to attend the conference and their long-term participation patterns, accordingly.

The association with a conference community and conference inclusiveness, can have a strong influence on scientists persistence in participating at the specific conference. The problem of the order-parameter persistence (first-passage time), is a well studied phenomenon in non-equilibrium statistical dynamics in condensed matter systems~\cite{majumdar1999persistence}. Persistence is defined as the probability that fluctuating variable does not change the sign until time $t$, and for many non-equilibrium systems this probability decays with time as a power-law~\cite{majumdar1999persistence}. Here we carry out the analysis of persistence of participation patterns of more than $100 000$ scientists at six national and international conferences of different sizes and from different fields of science. We study the probability of total and successive number of participations, as well as the distribution of time lags between two successive participations. We find that all three measured probabilities have a shape of a truncated power law, regardless of the conference size and degree of specialization. This indicates that the probability for a participant to attend the next meeting is not constant, but rather it grows/decays with a number of participations/non-participations. This observation is directly related to the strength of the association with the conference community. We propose a microscopic stochastic model which includes this influence of balance between the number of participations and non-participations, as well as the role of conference inclusiveness, on the probability to attend the conference next year. Results of our model show that the studied conferences have a relatively low inclusiveness, i.e. the probability for a scientist to participate in the next meeting after the first attendance. We also show that conference attendance is characterized by \textit{positive feedback}. The growth in the total number of participations results in a stronger attractiveness of the conference community to participants, and vice versa. Longevity of scientific career of publishing in scientific journals is also characterized by a power-law distribution with an exponential cut-off \cite{petersen2011}. Using the empirical analysis and stochastic model Petersen et al. \cite{petersen2011} have shown that longevity and past success of scientists lead to cumulative advantage in further development of their career. Although the distribution of career longevity and conference persistence have a similar behaviour, there is a significant difference of characteristic exponents, which indicates that a different mechanism underlie these two phenomena.

This paper is structured as follows: first, we perform empirical analysis of participation patterns for six conferences. We then propose and describe the model of conference participation dynamics. Finally, we perform numerical simulations and discuss some properties of the model, and estimate the values of parameters that correspond to empirical data.

\section*{Results}
\subsection*{Data set}
For our empirical analysis we use data for six conference series in different fields of science. We collected and filtered information about abstracts presented at the American Physical Society March Meeting (APSMM), American Physical Society April Meeting (APSAM), Society for Industrial and Applied Mathematics Annual Meetings (SIAM), Neural Information 
Processing Systems Conference (NIPS), International Conference on Supercomputing (ICS) and Annual International Conference on Research in Computational Molecular Biology (RECOMB). All these scientific meetings are held annually, but they differ in the topic, sizes, degree of specialisation, longevity and degree of localisation (national versus international). When it comes to the meeting size it can vary from a few dozens, like ICS and RECOMB, to several thousands of participants at APSMM. Some of these meetings are on highly focused topic, NIPS, while others are designed to cover the entire scientific fields, like APSMM, APSAM and SIAM. Four of these conferences (SIAM, NIPS, ICS and RECOMB) have an international character with venues all over the world, while APSMM and APSAM are annual conferences of American Physical Society which are always held in North American cities. APSMM, SIAM and APSAM are conferences with a long tradition, while first meetings of NIPS, ICS and RECOMB have been organized during late 80s and early 90s. Detailed information about conferences and data is given in Supporting Information (SI).

To be able to track participants at the conference over the years, we have labeled them based on name, affiliation and co-authors and performed author name disambiguation (see Methods for details). We are interested in studying the participation patterns of scientists starting from their first attendance at the conference series. Thus, for conferences for which the data are not available from their beginning (APSMM, APSAM and SIAM), we have filtered out the authors that may have attended the conference before the starting year in our dataset (see Methods for the details of our filtering procedure).

\subsection*{Empirical results}
For all scientists we have the information about the years of their appearance as authors in the book of abstracts of particular a conference series. The information about the list of authors who actually attended the conference is not available for the conferences considered in this paper. Hence, as a proxy for a conference participation in a given year, we use the appearance of a scientist as a co-author of at least one abstract in conference proceeding for that year. Not all authors that are mentioned in the book of abstracts have actually attended the conference, but one can argue that as co-authors they have actively contributed to the material presented and thus participate as a contributors in the conference~\cite{borghans2010}. 

\begin{figure}[ht]
\centering
\includegraphics[width=\linewidth]{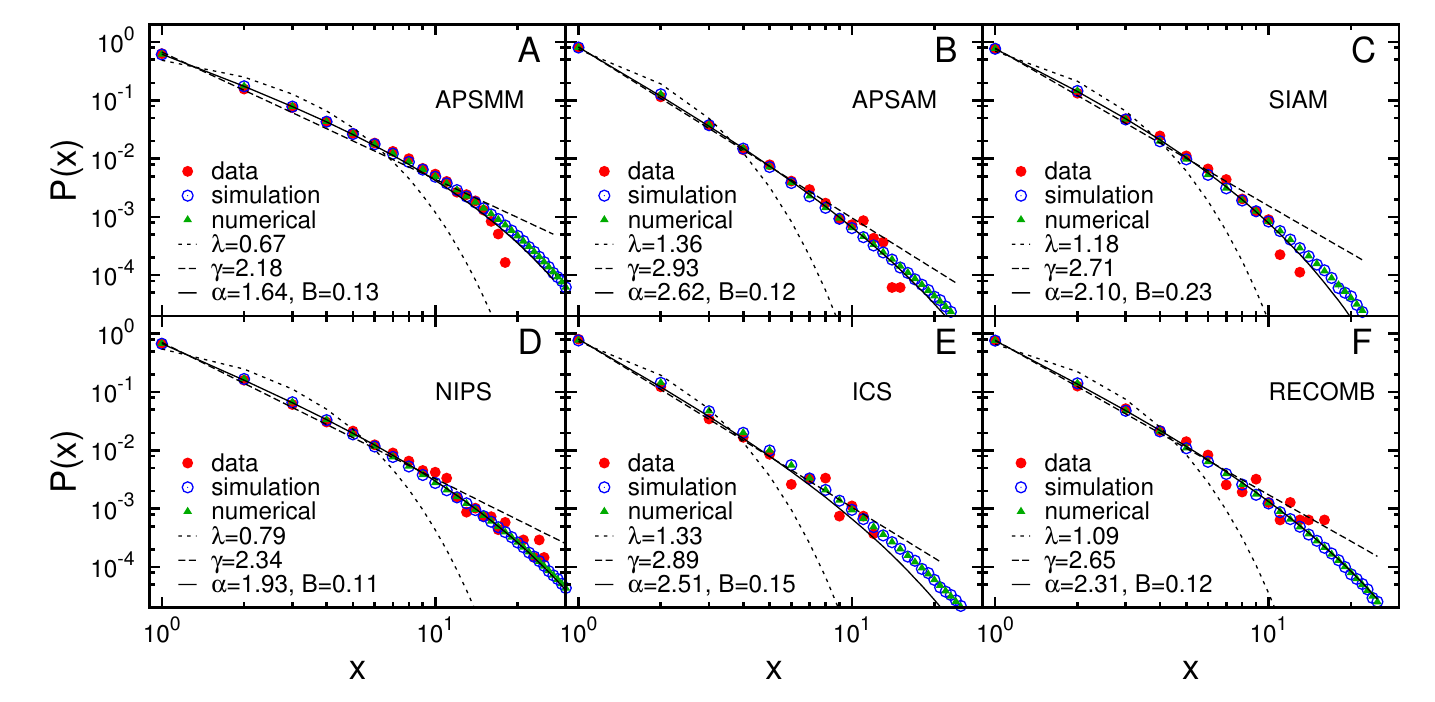}
\caption{\textbf{The total number of participations.} The probability distribution of the total number of participations obtained from the empirical data (red circles), simulations (blue circles) and numerical iterative algorithm (green triangles). The full line is the best fit to truncated power law, $x^{-\alpha}e^{-Bx}$, while the dashed and dash-dot line denote the best fit to power-law distribution, $x^{-\gamma}$ and exponential distribution, $e^{-\lambda x}$, respectively.}
\label{data_Nt}
\end{figure}

First we analyse \textit{the total number of author's participations} (the number of times an author has participated), $x$, at the given conference series. Figure~\ref{data_Nt}, shows the probability distribution of the total number of participations, $P(x)$, averaged over all participants, for each of the six analysed conferences. The comparison of the quality of fits between exponential, power-law and truncated power-law, Figure~\ref{data_Nt}, shows that all curves are very well represented by power law with exponential cut-off (see Methods), with the value of exponent $\alpha\in (1.6,2.7)$. The disparity in the total number of participations indicates that most scientists belong to the group of occasional participants, with more than half of all participants attending a particular conference only once. For instance, the percentage of all participants that attend the conference only once is the highest for APSAM and ICS, around $81\%$, and the lowest for APSMM and NIPS, $63\%$ and $68\%$ respectively. This observation indicates that communities of all these conferences have a relatively low inclusiveness. On the other hand, it is also clear that some of the participants are very regular, attending the conference (almost) every year. These participants form the group of regular attendees whose conference participation is mainly driven by social factors, i.e. their sense of \textit{association with the community}.

In the case of when the probability to attend a conference is constant or random, the expected distribution of total number of attendances is of exponential type. Thus, the power-law nature of the distribution of total participations strongly suggests that the probability of participation at some future conference increases with the number of previous participations. By participating frequently at a particular conference scientists not only expand, but also strengthen, their collaboration network which leads to their further engagement with the community.

\begin{figure}[th]
\centering
\includegraphics[width=\linewidth]{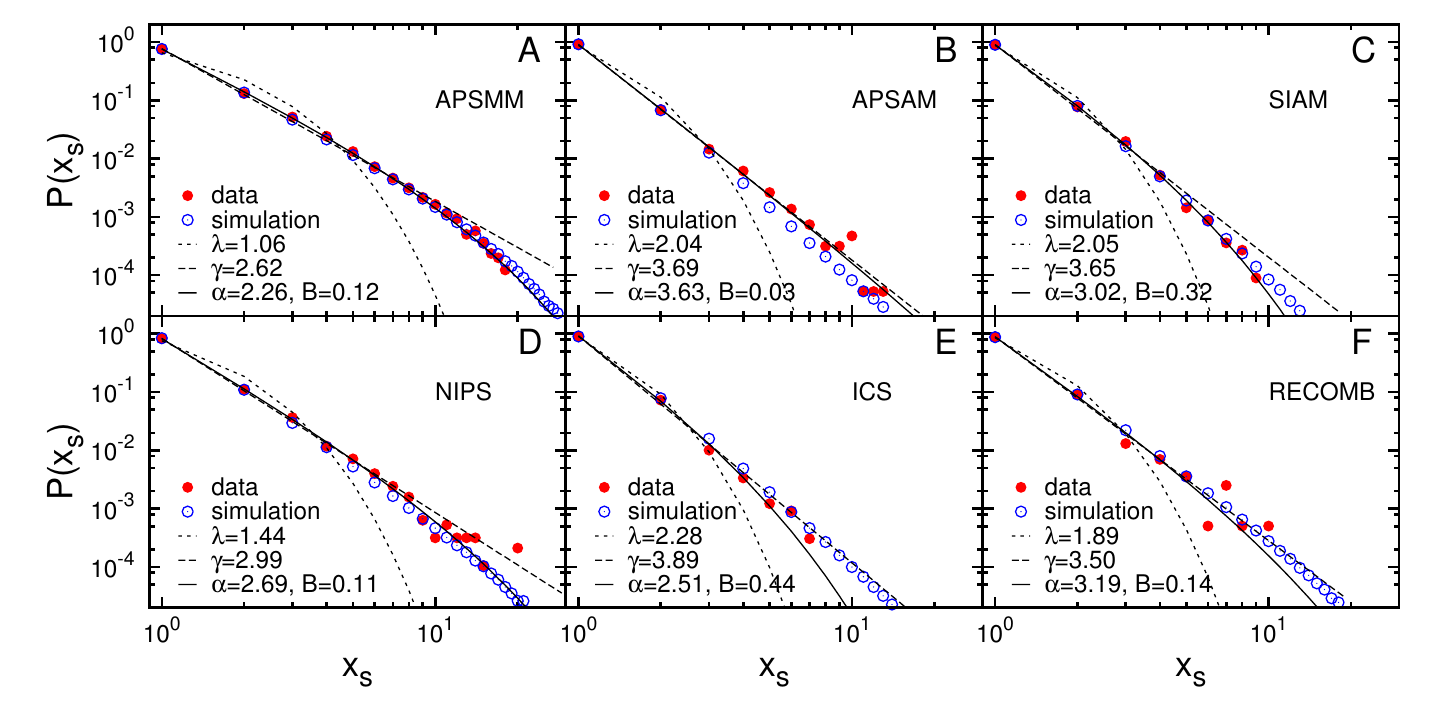}
\caption{\textbf{The number of successive participations.}The probability distribution of the number of successive participations, $x_{s}$, obtained from empirical data (red circles) and numerical simulations of the model (blue circles). The full, dashed and dash-dot line are the best fit to truncated power law, power-law and exponential function respectively.}
\label{data_Ns}
\end{figure}

\begin{figure}[th]
\centering
\includegraphics[width=\linewidth]{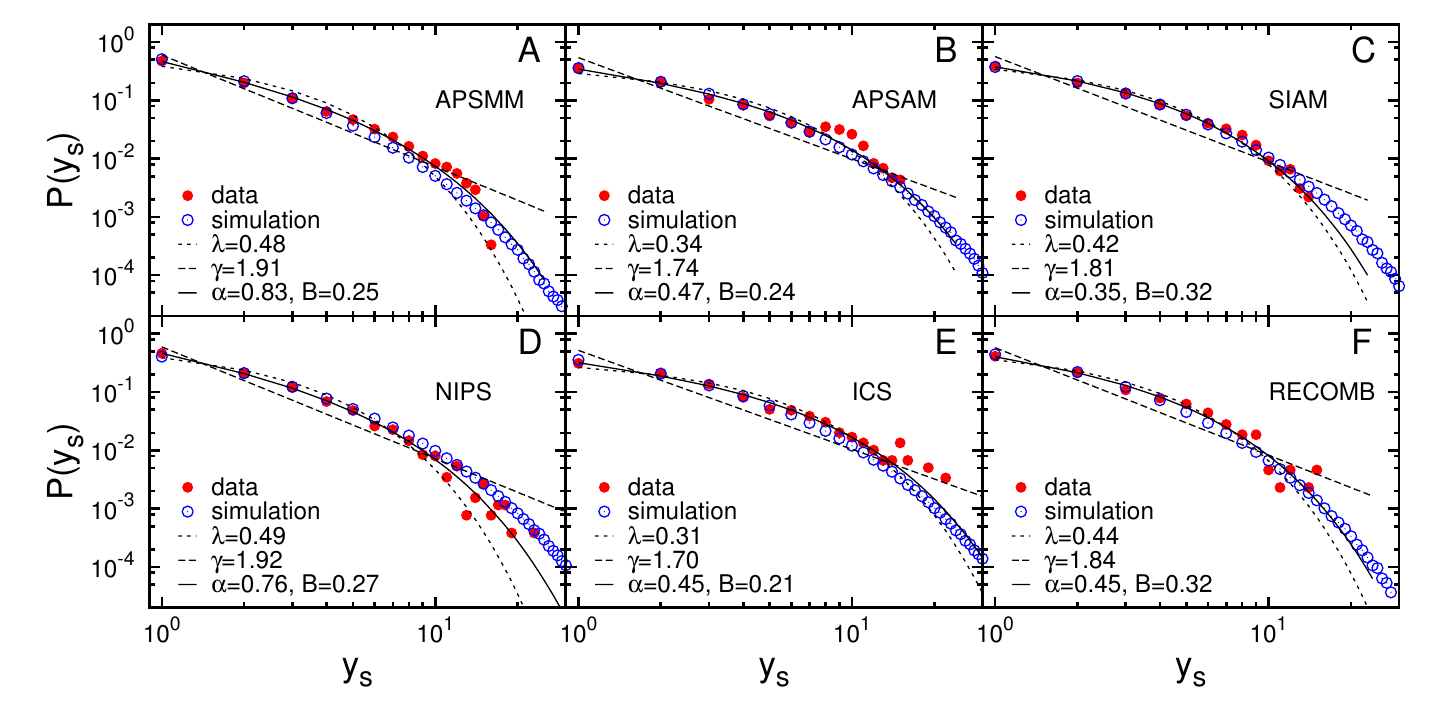}
\caption{\textbf{The time lag between the t\textsl{•}wo successive participations.}The probability distribution of the time lags between two consecutive conference participations $y_{s}$: empirical data (red circles) and numerical simulations data (red circles). The lines correspond to respective fits as in Figures~\ref{data_Nt} and \ref{data_Ns}.} 
\label{data_Ps}
\end{figure}

We further explore the participation patterns by analysing the number of successive participations (Figure~\ref{data_Ns}) and the time lag between two successive participations (Figure~\ref{data_Ps}). The distributions of these quantities also exhibit the truncated power-law behaviour (see Methods). The observed distributions of the number of successive participations, with exponent $2\leq\alpha\leq 4$, suggests that even frequent attendees make a pause in their participation, although these breaks are usually short, i.e. long breaks of five and more years occur with a low probability, Figure~\ref{data_Ps}. A long-period of non-participation results in fading of existing collaboration ties with the community while new ones are never formed. Due to this fading, the probability to attend the meeting decreases with total number of non-participations. This indicates that conference participation of most scientists takes place in a limited period of time with a relatively short and small number of breaks.

As it was shown in Ref.~\cite{petersen2011} the distribution of the journal career longevity exhibits a truncated power-law behaviour with cut-off around $10$ years. The exponential cut-off in the distribution of all three measures is a consequence of the two combined finite-size effects that influence the asymptotic behaviour, the finite life time of scientist's association with one community or her/his career in one field of research or in science in general \cite{petersen2011}, and limitations of used datasets.  This effect will be also observed in the distribution of conference participations. The end of a career inevitably results in a termination of participation in conferences and thus also the conference community membership. Also, used datasets have a relatively short time span (less than three decades), due to which they do not include scientists with long careers \cite{petersen2011}. Both of these effects affect the value of the exponential cut-off, which is lower in the case of conference participation, between $4$ and $9$ years, compared to the one observed for the career longevity.

\subsection*{Model}
The empirical results from six different series shown in the previous section indicate that the probability for a scientist to attend the next meeting of a conference series depends on the balance of previous participations and non-participations. Petersen et al. \cite{petersen2011} show that Matthew (\textit{rich get richer}) effect  is responsible for the career longevity in several competitive professions, including science. They argue that it becomes easier to move forward in the career with an increasing past success of an individual, and show, using their stochastic career progressive model, that this mechanism leads to a truncated power-law distribution of the career longevity. In their model, they assume that the stochastic process governing career progress is similar to Poisson process, where progress is made at any given step with the rate $g(x)\equiv 1-exp[-(x/_{x_{c}})^{\alpha}]$, where $1/x_{c}$ is a hazard rate corresponding to random career ending while the parameter $\alpha$ is the same as power-law exponent in the pdf of career longevity. Using this model for $\alpha<1$ they were able to obtain truncated power-law distributions for career duration in several professions.

The empirical results of conference participation patterns suggest that the probability for a scientist to participate in a conference is not constant or random, but that it rather grows with the number of participations. This is reflected in the increase of proportion of authors who are going to attend the conference next year with total number of previous conference attendance (see Figure 1 in SI). Higher number of participations of a scientist at the conference results in better connections with the community and thus higher probability that the author will participate in the following conference. But unlike career longevity, where the length of the waiting times between two successive steps in the career does not influence the progress rate, the probability for conference participation is strongly influenced by the number and length of pauses (Figure 2 in SI). The longer the scientists are absent from the community the weaker are their connections and lower are the probabilities to participate in the following events. For this reason and the fact that the pdf obtained from the model proposed in Ref. \cite{petersen2011} exhibits a truncated power-law only for the exponents $\alpha<1$ Petersen et al. model \cite{petersen2011} cannot be applied for modelling conference participation dynamics.

We propose a new stochastic model for conference attendance dynamics which can explain our empirical findings. Our model is based on a 2-bin generalized P\'olya process \cite{Drinea:2002:BBM:545381.545422,pemantle2007,RSA:RSA20261} 
and random termination time of a career. As opposed to the Petersen model where the progress rate depends only on the current position of scientist in his/her career, the 2-bin generalized P\'olya incorporates dependence on the balance between participations and non-participations. Let $x$ stands for the total number of participations at the conference, $y$ stands for the number of conferences an author has not participated
since she/he appeared at the conference for the first time and $t$ is the number of events held, $t=x+y$. All authors start with $x=1$ and $y=0$. According to our model, the probability that a scientist with $x$ total number of participations and $y$ number of non-participations will appear at the next conference is given by
\begin{equation}
 g(x,y)=\frac{x^p}{x^p+(y+y_0)^p}=\frac{z^{p}}{1+z^{p}}\ , 
 \label{g_formula}
\end{equation}
where $z=\frac{x}{y+y_{0}}$ measures the balance between participations and non-participations, parameter $p$ is the \textit{exponent} of the model, and $y_{0}$ determines the initial balance value. The probability that a scientist will not attend the next conference is equal to $1-g(x,y)$.   
Depending on the exponent $p$, the function $g$ can correspond to positive ($p>1$) or negative feedback ($p<1$) \cite{Drinea:2002:BBM:545381.545422}. When $p=1$ and $y_{0}=0$, the Eq.~\ref{g_formula} is equivalent to the equation for a P\'olya-Eggenberg problem \cite{johnson1977urn}. As we shall see in the following section, the value of the parameter $p$ for all conferences is larger than one, suggesting that the conference participation dynamics is characterized by the \textit{positive feedback}: scientists who participate in the conference frequently and make less and shorter pauses have a stronger association with the conference community and thus have a higher probability to participate in the following events. The value of the parameter $y_{0}$ determines the probability of a scientist to attend the next event after her/his first occurrence at the conference. According to our model this parameter is the same for all scientists attending one conference series, thus it can be interpreted as a measure of the conference community inclusiveness. 

\subsubsection*{Evolution equation}
The probability $P(x,t)$ for the author to have $x$ conference participations after $t$ conferences since his/her first participation is equal to the probability to attend the next conference $g(x-1,t-x)$ times the probability of already attending $x-1$ conferences at time $t-1$ plus the probability of skipping the next conference $1-g(x,t-1-x)$ times the probability of already attending 
$x$ conferences at time $t-1$:
\begin{equation}
 P(x,t)=\frac{(x-1)^p}{(x-1)^p+(t-x+y_0)^p}P(x-1,t-1)+\frac{(t-1-x+y_0)^p}{x^p+(t-1-x+y_0)^p}P(x,t-1). \label{P(x|t)_formula}
\end{equation}
The probability distribution $P(x)$ of the number of total conference attendances for a particular conference series is obtained by summing $P(x,t=T)$ over all possible $T$:
\begin{equation}
 P(x)=\sum_{T=1}^{\infty}P(x,t=T)P(T) \ , \label{P(x)_formula}
\end{equation}
where $T$ denotes the duration of a scientist's membership in the community. In our case, we assume that the duration of a scientist's membership in a conference community can be terminated at any year after his/her first appearance with probability $H$, which gives the distribution of time intervals      
\begin{equation}
 P(T)=H(1-H)^{T-1}. \label{P(T)_formula}
\end{equation}

\subsection*{Numerical simulation results}
Since the analytical solution of Eq.~\ref{P(x)_formula} cannot be obtained, we estimate the model parameters $y_0$, $H$ and $p$ using numerical simulations (see Methods). The best estimates of the model parameters for each of the six conferences are given in SI Table 7.  As shown in Figures~\ref{data_Nt}, \ref{data_Ns} and \ref{data_Ps}, the model with the properly chosen parameters nicely reproduces the behaviour of participants at six conferences, for all three measured quantities.

For all six conferences the estimated value of parameter $p$ is greater than $1$, which suggests that the positive feedback mechanism underlies the conference participation dynamics. This means that the probability for a scientist to attend the next year event grows superlinearly with the balance between the number of participations and pauses ($z$). The value of the parameter $y_{0}$ together with the value of $p$ determines the probability for a scientist to participate in the conference next year after his/her first participation, i.e. the initial inclusiveness of the conference community. Table S8 in SI shows the estimated value of the initial inclusiveness for all six conferences. The APSMM has the highest probability, around $25\%$, for newcomers to attend the conference next year, while APSAP has the lowest, $9\%$. One could assume that the size and diversity of topics of a conference have an essential influence on conference inclusiveness, but according to our results this is not the case. The ordering of the conferences according to size, Table S7 in SI, and their initial inclusiveness do not correlate. APSAM is the second largest conference but has the lowest inclusiveness, while the RECOMB as the smallest conference is ranked as third and has the inclusiveness of $15\%$. Further, it follows from our results that the diversity of topics covered by the conference does not have a significant effect on the return probability of newcomers. Although the first ranked conference according to inclusiveness, APSMM, covers the widest range of topics among considered conferences, the APSAM and SIAM, which are also considered general conferences, have a lower inclusiveness than NIPS and RECOMB. This suggests that the conference inclusiveness is influenced by some other factors, which are not related to the size, degree of specialisation or localisation (national and international), but rather to social structure and openness of the conference community toward newcomers.

We solve Eq.~\ref{P(x)_formula} numerically using an iterative method (see SI for more details) and compare it with simulation results. Figure~\ref{data_Nt} shows an excellent matching between results obtained using the iterative algorithm and numerical simulations for the estimated values of parameters.

\section*{Discussion and conclusion}

The goal of this paper has been to investigate the conference participation patterns and propose a simple stochastic model of conference participation dynamics. The motivation behind this is to better understand the mechanisms that underlie the repeated participation in the same conference series and explore whether the conference series topic, size, degree of specialisation, longevity and degree of localisation (national and international) influence the participation probability and inclusiveness of the specific community. Our study is based on empirical analysis and modelling of authors participation at six different conference series in the last three decades: APSMM, APSAM, SIAM, NISP, ICS and RECOMB. We note here that it would be important to verify our findings with the data from other conferences.   

The set of  considered  conferences is very heterogeneous. Although they differ in size, topic and topic diversity, national structure of participants and conference longevity, they are characterized with similar participation patterns. The distributions of the total number of participations for all six conferences exhibit the same, truncated power-law, behaviour with values of exponent $\alpha$ between $1.6$ and $2.7$. A similar behaviour is also observed for the distributions of the number of successive participations and the duration of pauses between them. The observed statistical evidence strongly imply that the dynamics of conference participation is governed by universal forces which are independent of the specific conference features or the scientific field. This and the fact that conferences often have a stable core of attending participants \cite{vandijk2006} suggests that these have social origins and that social factors, such as the association with a conference community and its inclusiveness, strongly influence the probability for a scientist to attend the future meetings and their participation patterns at the specific conference series, accordingly.

The observed truncated power-law behaviour of the distributions of participations indicates that the probability for a scientist to participate in the next year conference is growing(decreasing) with the balance between the number of participations and pauses. To further explore this we proposed a stochastic model based on 2-bin generalized P\'olya process which incorporates the dependence on the ratio between number of participations an pauses. Our model shows that the positive feedback mechanism underlies the conference participation dynamics. The probability for a scientist to attend a conference grows superlineary with the number of participations, while the frequent pauses have the opposite effect. The scientists who are able to overcome the initial obstacles and create social ties with the conference community by frequent participation at the beginning have a higher probability to attend the conference in the following years. A frequent participation strengthens the scientist's association with a conference community which further increases the probability for future participations. On the other hand, scientists with a small number of initial participations have a low probability to participate in the following conference, thus small number of participations, and eventually stop attending the conference. The initial inclusiveness of the specific conference community has the main influence on early participation patterns. As we showed, this inclusiveness does not depend on the size, degree of specialisation or topic of the conference, but rather on the openness of the community toward newcomers.
       
Our analysis indicates that social factors, such as the association with the community and the community inclusiveness are the main driving forces of conference participation dynamics. In general the community/group cohesion and the ability to attract and retain newcomers and other members influence the dynamics of their participation in group activities \cite{friedkin2004}. On the other hand, a member's engagement in group activities strengthens ties to other group/community members, and contributes to the creation of the bonding capital, while the ties of non-attendees dissolve and weaken with time \cite{sessions2010}. Conference communities are just one example of these systems, thus we expect to observe the similar group participation patterns in other types of social communities, both online and offline. Further investigations and studies of other social systems will reveal and characterize the connection between a social network structure and group inclusiveness, and participation dynamics in group activities.        

\section*{Methods}

\noindent
\textbf{Data filtering:}
Identification of the different authors may involve a few issues. On one hand, an author may use different spelling variants to sign his first 
and middle name. On the other hand, the author's name may be related to several different authors, thus using only the initials of the last name and first name increases  additionally error rates in disambiguating the author names. In our data sets, data from NIPS and RECOMB conferences did not require additional cleaning, while for the SIAM and ICS data, we have used python fuzzy partial string matching of 
author's first and middle names, which gave a high accuracy. For APSMM and APSAM conferences, where data are highly heterogeneous, we have used a method described in \cite{wu2013sci} to disambiguate the author names. This method considers pairs of names that match on last name and first name initials. Then it groups the authors based on their affiliation 
and co-authors. Because the same affiliation could be formatted differently, the two affiliations were considered the same if their fuzzy token set ratio was higher than $50\%$.

The sources and detailed description of the data are given in SI Tables 1, 2 and 3. For NIPS, ICS and RECOMB, we have complete data from their very beginning. Remaining data sets required filtering out the authors with a high probability of attending conference before the starting year in our dataset, $Y_0$. Therefore, for APSMM, APSAM and SIAM we have isolated authors with the first recorded year of conference attendance, smaller than $Y_0+\langle \tau \rangle$, where $\langle \tau \rangle$ is the average waiting time between a
consecutive conference attendance for all the authors who took part at the conference during the $[Y_0,Y_f]$ period. This way we excluded between $10\%$ (APSMM and SIAM) and $25\%$ (APSAM) authors from our analysis.

\noindent
\textbf{Functional fits:}
We have used the maximum-likelihood fitting method~\cite{doi:10.1137/070710111} to fit three different functions to the probability distributions of the total number of participations, the number of and the time lags between two successive participations: exponential function $e^{-\lambda x}$, power-law function $x^{-\gamma}$ and truncated power-law $x^{-\alpha}e^{-Bx}$. It follows from the comparison of fits of these three functions to empirical data that the truncated power-law is the best fit for the probability distribution of all three measured quantities, see Figures ~\ref{data_Nt}, \ref{data_Ns} and \ref{data_Ps}. In order to compare these three fits we calculate the log likelihood ratio, $\mathcal{R}$, and $\pi$-value (see Ref.~\cite{doi:10.1137/070710111}) which compares the fits to the power-law with exponential cut-off with the pure power-law for the distribution of total number of participations (SI Table 4) and the number of successive participations (SI Table 5). In the case of nested distributions, the negative value of $\mathcal{R}$ indicates that the larger family of distributions, in this case the truncated power-law, is a superior model. When the value of $\mathcal{R}$ tends to $0$, one can use $\pi$-value. The small $\pi$-value  suggests that the smaller family of distributions, in this case power-law, can be ruled-out. Both the log likelihood ratio and the $\pi$-value indicate that the truncated power-law is a superior model compared to pure power-law for both distributions. A similar procedure can be applied for the comparison between truncated power-law and exponential fits, but since from the visual inspection it is clear that the distributions do not follow the exponential fits, we have omitted these results. The comparison between exponential and the power-law with exponential cut-off fit, given in SI Table 6, indicates that the power-law distribution with exponential cut-off fit is better than exponential fit for the distribution of the time lags. For all six conferences, the power-law with exponential cut-off distribution gives the best fit for all three empirical distributions.

\noindent
\textbf{Parameter estimation:}
We simulate the model for $N=100000$ different authors. Starting from $x=1$ and $y=0$ at $t=1$, an author will appear at the next conference with probability $g(x,y)$ or skip it with the probability $1-g(x,y)$. The author can terminate his/her membership in the community at each time step with the probability $H$. In order to estimate the values of parameters $p$, $y_{0}$ and $H$,  we calculate the distribution of total number of attendances $x$, from the simulations and compare it to the empirical distribution using Kullback-Leibler 
Distance \cite{burnham2002model}. We perform the simulations for several different sets of parameter $(y_0,H,p)$ to determine which combination of parameter values makes the model optimally close to the empirical data. For each parameter set the results are averaged across $100$ simulations.

\section*{Acknowledgements}

J.S. and M.M.D. gratefully acknowledge financial support by the Ministry of Education, Science, and Technological Development of the Republic of Serbia under project ON171017, by the European Commission under H2020 project VI-SEEM, and from the the European Community’s COST Action TU1305 "Social Networks and travel behavior". Numerical simulations were run on the PARADOX supercomputing facility at the Scientific Computing Laboratory of the Institute of Physics Belgrade, supported in part by the Ministry of Education, Science, and Technological Development of the Republic of Serbia under project ON171017. T.K. and M.M.D. acknowledge support from  from the European Community’s COST Action TD1210 KNOWeSCAPE. 

\section*{Author contributions statement}
J.S., A.C., T.K. and M.M.D. designed the research and participated in the writing of the manuscript. J.S. collected the empirical data, developed the model in silico and performed the numerical simulations. 

\section*{Additional information}
\textbf{Competing financial interests:} The authors declare no competing financial interests. 

\onecolumngrid
\newpage

\numberwithin{table}{section}
\numberwithin{figure}{section}

\begin{center}
\begin{large}
\textbf{Supporting Information}
\end{large}
\end{center}

\section{Data}
\subsection{Conference description}
The American Physical Society March Meeting (APSMM) is the world's largest condensed matter physics conference with more than 70 years history. It is organized annually 
at various locations in The United States. The conference attracts researchers from research institutions, universities, and industry from all around the world.

The APS April Meeting (APSAM) conference is dedicated to the topics from the astrophysics, gravitational physics, nuclear physics, and particle physics. Likewise March Meeting, the
conference takes place at various locations in The United States each year.

The Annual Meeting of the Society for Industrial and Applied Mathematics (SIAM) has been held since 1984 at various locations in The North America. Topics covered at the SIAM conferences include applied and computational mathematics and applications. 

The Neural Information Processing Systems (NIPS) Conference has been held since 1988 at various locations in The United States, Canada and Spain. Neural information processing intends to emerge fields such as machine learning, statistics, applied mathematics and physics. The acceptance rate is about $50\%$.

The aim of The International Conference on Supercomputing (ICS) is to promote an international forum for the presentation and discussion on the various aspects of high-performance computing systems. The ICS conference has been sponsored by The Association for Computing Machinery (ACM). The conference is organized annually since 1988 at various locations.
The overall acceptance rate is $26\%$.

The Annual International Conference on Research in Computational Molecular Biology (RECOMB) has been held since 1997 at various locations. At RECOMB emphasis is placed on 
connecting the biological, computational, and statistical sciences. The overall acceptance rate is $20\%$.

The list of links to the conference data and proceedings is given in  Table~\ref{linkW}, while the Table~\ref{sizecon} lists the sizes of all six conferences for all years covered in the data set. The number of participants is calculated after the name disambiguation.

\subsection{Data description}
\begin{table}[h]
\centering
  \begin{tabular}{|c|c|}
      \hline
                    Conference & Link to the conference data set \\  \hline \hline
  APSMM & http://www.aps.org/meetings/baps/ \\ \hline
  APSAM & http://www.aps.org/meetings/baps/ \\ \hline
  SIAM & http://www.siam.org/meetings/archives.php$\#$AN \\ \hline
  NIPS & http://papers.nips.cc/ \\ \hline
  ICS & http://dl.acm.org/event.cfm?id=RE215$\&$tab=pubs \\ \hline
  RECOMB & http://www.recomb.org/history \\ \hline
  \end{tabular}
\caption{Pages on the web from which we downloaded conference data.}
\label{linkW}
\end{table}

\begin{table}[h]
\centering
\begin{threeparttable}
\begin{tabular}{|c|c|c|c|}
 \hline \hline
{\it Conference} & {\it $Y_0$} & {\it $Y_f$} & {\it Number of participants}\\
\hline \hline
APSMM & 1994 & 2014 & 78544 \\ \hline
APSAM \tnote{*}& 1994 & 2014 & 16264 \\ \hline
SIAM \tnote{**}& 1994 & 2014 & 8879 \\ \hline
NIPS & 1988 & 2014 & 6902 \\ \hline
ICS & 1988 & 2014 & 2504 \\ \hline
RECOMB & 1997 & 2014 & 1564 \\ \hline
     \end{tabular}
     \begin{tablenotes}
       \item[*] Data are not available for 1999.
       \item[**] Data are not available for 2007 and 2011.
     \end{tablenotes}
  \end{threeparttable}
\caption{Summary of the conference data. Columns 2 and 3 indicate for each conference the year in which data we have collected begin ($Y_0$) and end ($Y_f$). 
The total number of different participants at the conference during that period of time is given in column 4.} \label{conferences}
\end{table}

\begin{table}[h]
\centering
  \begin{tabular}{|c|c|c|c|c|c|c|}
      \hline
                    & APSMM & APSAM & SIAM & NIPS & ICS & RECOMB \\  \hline \hline
  1988&-&-&-&214&132&- \\ \hline
  1989&-&-&-&205&121&- \\ \hline
  1990&-&-&-&297&123&- \\ \hline
  1991&-&-&-&302&116&- \\ \hline
  1992&-&-&-&270&112&- \\ \hline
  1993&-&-&-&301&114&- \\ \hline
  1994&9660&3309&540&270&114&- \\ \hline
  1995&9897&1947&425&292&144&- \\ \hline
  1996&9991&2356&279&289&127&- \\ \hline
  1997&9191&3388&579&289&109&111 \\ \hline
  1998&10924&2301&456&298&158&120 \\ \hline
  1999&20426&-&367&296&172&121 \\ \hline
  2000&10816&1744&403&307&105&150 \\ \hline
  2001&12401&1818&823&396&146&101 \\ \hline
  2002&11944&2446&1115&432&118&98 \\ \hline
  2003&13548&2127&642&469&103&95 \\ \hline
  2004&14595&1668&767&492&102&136 \\ \hline
  2005&14673&1140&792&515&165&141 \\ \hline
  2006&16484&1008&945&479&124&154 \\ \hline
  2007&16655&943&-&530&96&123 \\ \hline
  2008&16441&1473&1053&633&132&142 \\ \hline
  2009&16775&1630&1054&654&242&127 \\ \hline
  2010&17790&1342&1166&733&127&157 \\ \hline
  2011&18368&1088&-&746&171&167 \\ \hline
  2012&22343&1480&1223&938&133&148 \\ \hline
  2013&21510&1430&1205&884&210&125 \\ \hline
  2014&22789&1704&1407&1064&147&137 \\ \hline
  \end{tabular}
\caption{The number of participants at the conference per year and the total number of participants.}
\label{sizecon}
\end{table}

\section{Functional fits}
We use maximum-likelihood to estimate the parameters of three different functions, exponential, power-law and power-law with an exponential cutoff for the distributions 
of total number of participations, the number of and the time lag between two successive participations. Further on, we calculate the log-likelihood ratio, $\mathcal{R}$, 
and $\pi$-value~\cite{doi:10.1137/070710111} between different fits in order to estimate which of the three different functional forms the best fits with the empirical observations. 
The Tables~\ref{R_Nt} and~\ref{R_Ns} show $\mathcal{R}$, and $\pi$-value calculated for the comparison between truncated power-law and pure power-law for total and successive 
number of participations, while Table~\ref{R_Ps} shows the comparison between fits of exponential and truncated power-law to the distribution of time lags. These results and visual 
inspection show that the power-law with an exponential cutoff is the best fit for all three empirical distributions, and for all six conferences.

\begin{table}[h]
 \begin{center}
  \begin{tabular}{|c|c|c|}
      \hline
                    & $\mathcal{R}$ & $\pi$     \\  \hline \hline
      APSMM & -1758.44 & 0.0  \\ \hline
      APSAM & -36.89 & 0.0  \\ \hline
      SIAM & -75.26 & 0.0  \\ \hline
      NIPS & -76.64 & 0.0  \\ \hline
      ICS & -8.54 & 3.60e-05 \\ \hline
      RECOMB & -7.22 & 1.45e-04 \\ \hline
  \end{tabular}

 \end{center}
\caption{Log likelihood ratio $\mathcal{R}$ and the $\pi$-value compare the fit to the power-law with the fit to the power-law with an exponential cutoff for the probability 
distribution of number of conferences at which each author appears.}
\label{R_Nt}
\end{table}

\begin{table}[h]
 \begin{center}
  \begin{tabular}{|c|c|c|}
      \hline
                    & $\mathcal{R}$ & $\pi$     \\  \hline \hline
      APSMM & -554.05 & 0.0  \\ \hline
      APSAM & -0.77 & 0.21  \\ \hline
      SIAM & -17.98 & 2.01e-09  \\ \hline
      NIPS & -17.52 & 3.24e-09  \\ \hline
      ICS & -4.99 & 1.57e-03 \\ \hline
      RECOMB & -1.48 & 0.09 \\ \hline
  \end{tabular}
 \end{center}
\caption{Log likelihood ratio $\mathcal{R}$ and the $\pi$-value compare the fit to the power-law with the fit to the power-law with an exponential cutoff for the probability 
distribution of the number of successive participations at the conference.}
\label{R_Ns}
\end{table}

\begin{table}[h]
 \begin{center}
  \begin{tabular}{|c|c|c|}
      \hline
                    & $\mathcal{R}$ & $\pi$     \\  \hline \hline
      APSMM & -756.91 & 0.0  \\ \hline
      APSAM & -34.59 & 1.11e-16  \\ \hline
      SIAM & -11.54 & 1.55e-06  \\ \hline
      NIPS & -58.22 & 0.0  \\ \hline
      ICS & -7.64 & 9.24e-05 \\ \hline
      RECOMB & -3.60 & 7.27e-03 \\ \hline
  \end{tabular}
 \end{center}
\caption{Log likelihood ratio $\mathcal{R}$ and the $\pi$-value compare the fit to the exponential with the fit to the power-law with an exponential cutoff for the probability 
distribution of the time lag between two consecutive conference participations.}
\label{R_Ps}
\end{table}

\section{Model of conference attendance dynamics}
\subsection{Participation probability}
\begin{figure}[h]
 \begin{center}
  \includegraphics[scale=0.4]{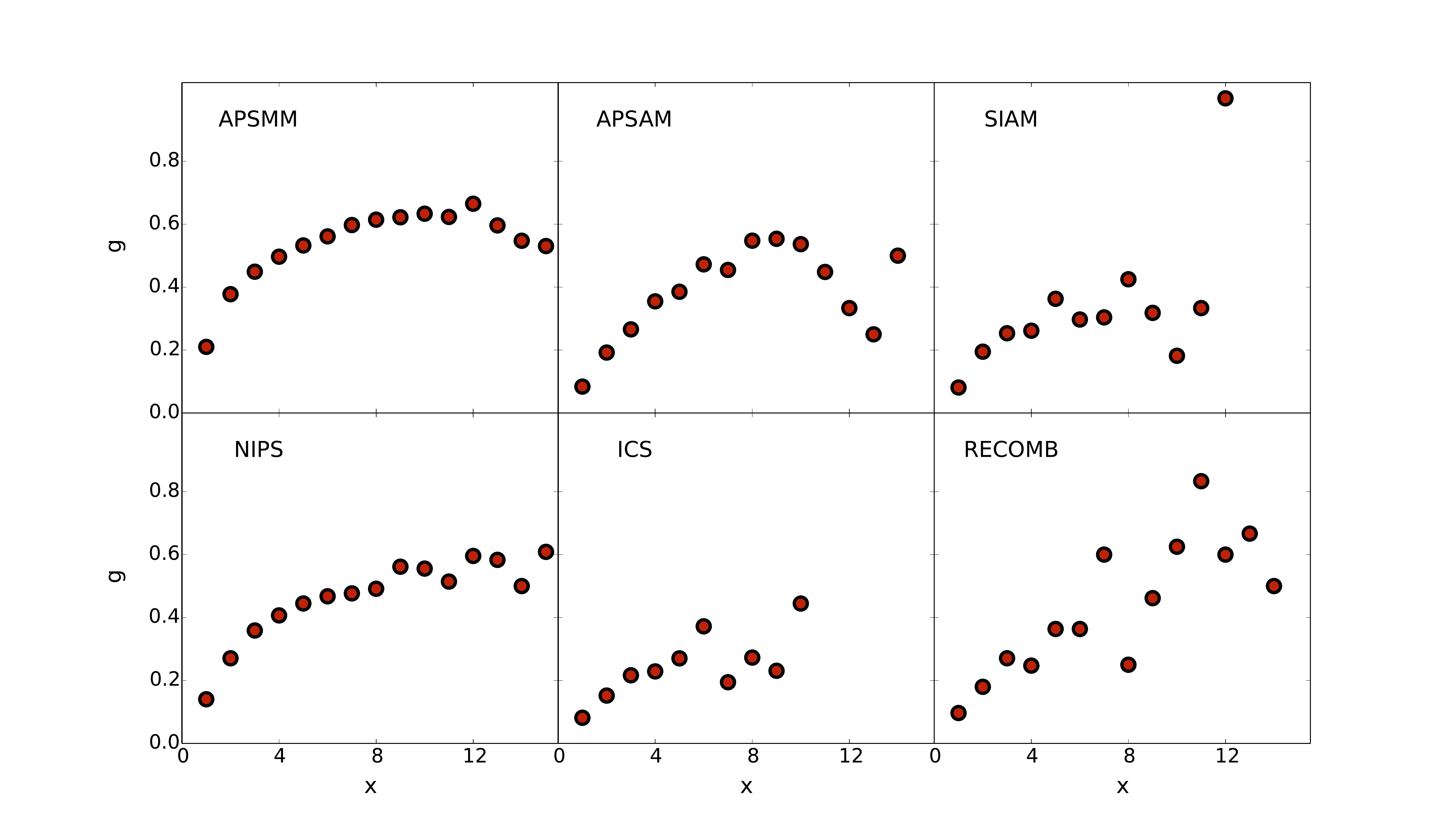}
 \end{center}
\caption{Proportion of conference participants $g$ with $x$ conference attendances who are going to attend the conference next year.} 
\label{data_G}
\end{figure}

\begin{figure}[h]
 \begin{center}
  \includegraphics[scale=0.4]{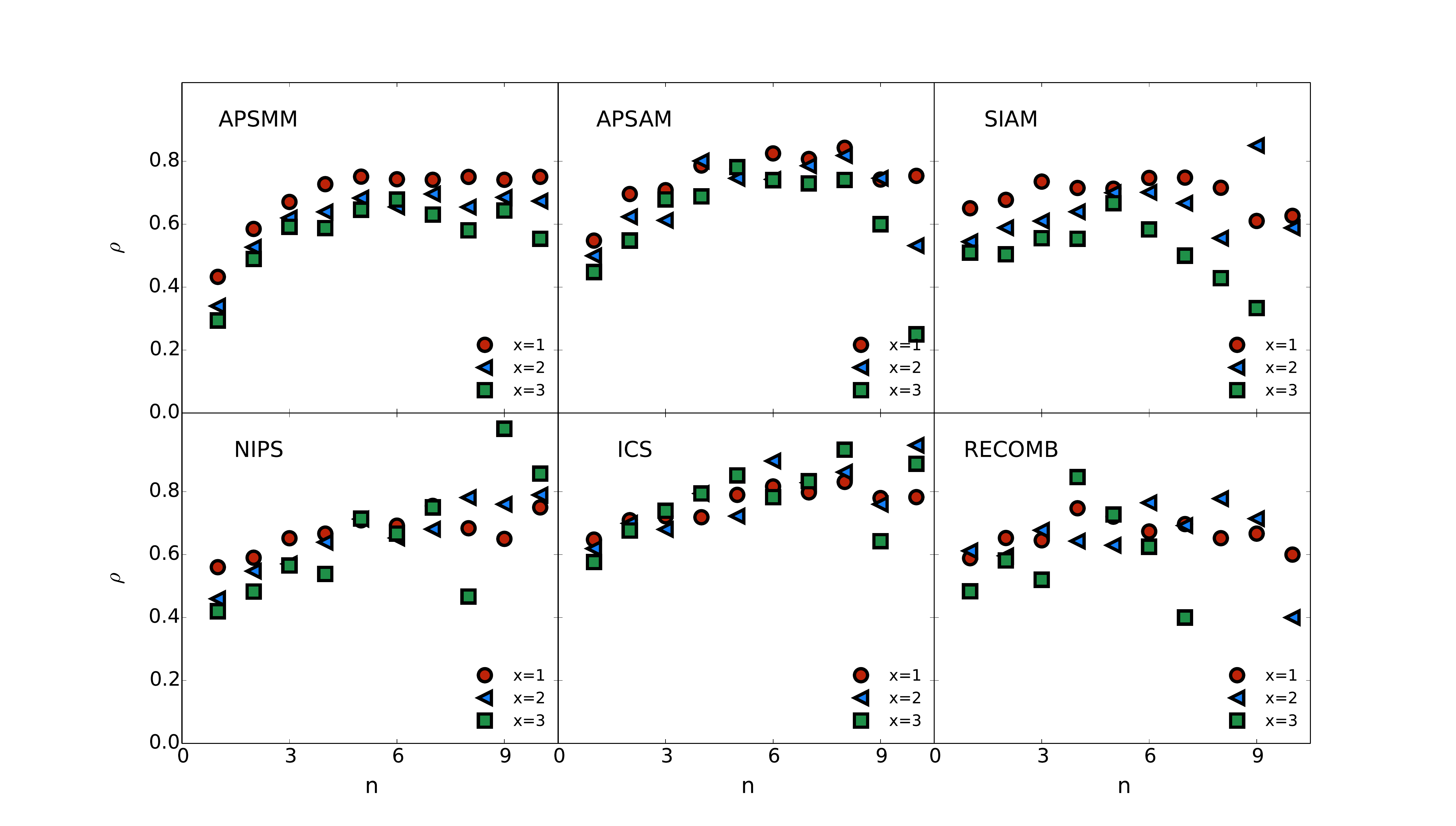}
 \end{center}
\caption{Proportion of conference participants $\rho$ with $n$ missed conferences after $x$-th conference attendance who are going to skip the conference next year, but will
take part at some future conference from the observation period.} 
\label{data_B}
\end{figure}

Figure~\ref{data_G} shows how the probability to attend the next meeting is changing with the number of previous attendances, calculated from the empirical data. We see that for 
all six conferences this probability grows for a small number of attendances. The saturation or decrease in the probability for a large number of previous participations, observed for 
some conferences, occurs due to a small number of observations for the large number of participations/length of pauses.\\   
Figure~\ref{data_B} shows how the probability to not attend the next meeting, $\rho=1-g$, increases with the number of non-participations, $n$, for the fixed number of previous 
participations $x$. We see that this probability is higher for smaller $x$ and the same value of $n$.\\
\subsection{Parameter estimation}
The Table~\ref{optparam} shows the optimal parameter values of the model for the six different conferences. In Table~\ref{comp} we show the estimated values of conference 
inclusiveness, $g(1,0)$, and the order of the conferences according to this value and the value of exponent $\alpha$.  
\begin{table}[h]
 \begin{center}
  \begin{tabular}{|c|c|c|c|}
      \hline
                    & $y_0$ & $H$ & $p$     \\  \hline \hline
      APSMM & 2 & 0.165 & 1.550  \\ \hline
      APSAM & 4 & 0.135 & 1.700  \\ \hline
      SIAM & 4 & 0.155 & 1.525  \\ \hline
      NIPS & 3 & 0.130 & 1.525  \\ \hline
      ICS & 4 & 0.135 & 1.575 \\ \hline
      RECOMB & 3 & 0.175 & 1.675 \\ \hline
  \end{tabular}
 \end{center}
\caption{The optimal parameter values of the model for each conference.}
\label{optparam}
\end{table}

\begin{table}[h]
 \begin{center}
  \begin{tabular}{|c|c|c|c|c|}
      \hline
                    & order & $1-g(1,0)$ & order & $\alpha$     \\  \hline \hline
      APSMM & 1 & 0.2546 & 1 & 1.64 \\ \hline
      APSAM & 6 & 0.0865 & 6 & 2.62 \\ \hline
      SIAM & 4 & 0.1077 & 3 & 2.10 \\ \hline
      NIPS & 2 & 0.1577 & 2 & 1.93 \\ \hline
      ICS & 5 & 0.1012 & 5 & 2.51 \\ \hline
      RECOMB & 3 & 0.137 & 4 & 2.31 \\ \hline
  \end{tabular}
 \end{center}
\caption{Stagnancy rate $1-g(1,0)$ at $t=1$ for each conference and exponent $\alpha$ of power-law with an exponential cutoff distribution fit with the corresponding conference order.}
\label{comp}
\end{table}

\subsection{Iterative method}
\begin{figure}[h]
 \begin{center}
  \includegraphics[scale=0.4]{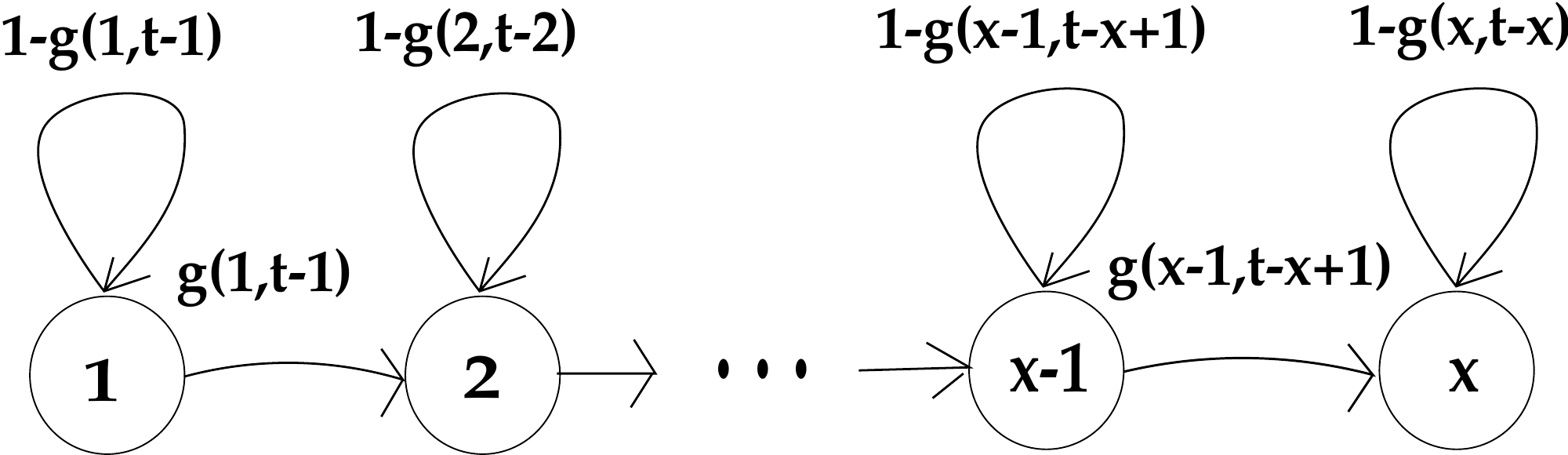}
 \end{center}
\caption{Scheme of the evolution of the process through one time step.} 
\label{scheme}
\end{figure}

The model evolution equations cannot be solved analytically, thus we use a numerical simulation and an iterative method. Here we explain the iterative method in details. 
Figure~\ref{scheme} is a schematic representation of the evolution process, which is a type of a Markovian process between states. Each state represents the number of 
participations. At each time step, the scientist can either attend a conference, with probability $g(x,t-x)$, and move one state right and increase the total number of 
participations, or not, and thus stay at the same state.  

To mathematically describe this evolution process we construct the transition probability matrices $M(t)$ of sizes $t \times t$, with elements
\begin{equation*}
M_{i,j}(t)= \left\{
  \begin{array}{l l l}
    1-g(i,(t-1)-i), & j=i \text{ and } i<t,\\
    g(i,(t-1)-i), & j=i+1 \text{ and } i<t,\\
    0, & \text{otherwise}.
  \end{array} \right.
\end{equation*}
for $t\geq 2$ and $M(t=1)=\text{I}$ at $t=1$.
The diagonal elements $M_{i,i}(t)$ define the probability that a participant who has $i$ attendances on $t-1$ conferences, does not attend conference at time $t$, 
while $M_{i,i+1}(t)$ represents the probability for the transition $i \to i+1$. We assume that the termination time of a conference career $T$ is the same for all participants and observe matrix
\begin{equation}
 \mathcal{M}=M'(1)M'(2) \dots M'(T-1)M(T)
\end{equation}
where $M'(t)$ is the matrix $M(t)$ expanded to the size $T \times T$ by adding $T-t$ zero rows and columns. The resulting matrix $\mathcal{M}$ has non-zero elements at the first 
row, and other elements are $0$. Each element $\mathcal{M}_{1,i}$ of the matrix $\mathcal{M}$ is the sum of all the possible combinations of attended and skipped conferences that 
result in $i$ total participations at time $T$. Otherwise stated, the ratio of authors who attended $i$ conferences is given by $\mathcal{M}_{1,i}$.\\

Based on this consideration, we next examine the probability distribution of the total number of participations when the termination of attendance occurs at random with some constant 
probability $H$. We generate matrices $\mathcal{M}(t)$:
\begin{itemize}
 \item $t=1$,\ $\mathcal{M}(1)=HM(1)$;
 \item $t=2$,\ $\mathcal{M}(2)=\left[\frac{1-H}{H}M'(1)\right]\left[HM(2)\right]$;
 \item $t=3$,\ $\mathcal{M}(3)=\left[\frac{1-H}{H}\mathcal{M}'(2)\right]\left[HM(3)\right]$;
 \item $t=T_{max}$,\ $\mathcal{M}(T_{max})=\left[\frac{1-H}{H}\mathcal{M}'(T_{max}-1)\right]\left[HM(T_{max})\right]$;
\end{itemize}
where $\mathcal{M}'(t)$ is the matrix $\mathcal{M}(t)$ expanded to the size $t+1 \times t+1$ by adding a zero row and column.
Each of elements $\mathcal{M}_{1,i}(t)$ of the matrix $\mathcal{M}(t)$ gives a ratio of participants that terminated their conference career at time $t$ with $i$ participations. 
We can choose $T_{max}$ to stop the calculation when $\sum_{i=1}^{T_max}\mathcal{M}_{1,i}(T_max) \to 0$.
Then, probability distribution of the total number of participations $P(x)$ is obtained by summing  over all possible career termination times
\begin{equation}
 P(x)=\sum_{t=1}^{T_{max}}\mathcal{M}_{1,x}(t).
\end{equation}

\end{document}